\documentstyle[12pt]{article}

\title{Comment on the "Density profiles in the theory of  condensation"
(Physica A, 226 (1996) 117-136).
Stationary profiles of the condensation}
\author{V.Kurasov}

\date{St.Petersburg State University}

\begin{document}
\maketitle

The theory of the density profiles of condensation \cite{PhysicaA} leads
to the radically new approach to the kinetics of the condensation process.
The formalism presented in \cite{PhysicaA} is complete but one can make
some simplifications in order to see how this theory transfers into the
old theory with the integral style of the account of the vapor consumption.
We shall call this old manner of account as the additive theory (AT).

The AT works in some conditions where the direct suppositions of this
theory (the homogeneous character of the vapor consumption from the whole
volume, the homogeneous profiles of the vapor density
around the droplet) aren't valid. The reason is rather simple and is based on
the
following important note {\it For the validity of the AT it is necessary
that the main substance  is accumulated by the part of the profile where
the decrease of the supersaturation is less or equal than $1/\Gamma$.
The concrete form of spectrum isn't important. It is necessary to know
the integral over the whole distances and it is given due  to the conservation
of the substance by the number of the molecules inside the droplet. }

The main numerical effect of the theory with the profiles of the condensation
(PT) is attained due to the gap of the density near the droplet.
One can see at least two opportunities to come to the additional regime:
\begin{itemize}
\item
There is no gap near the droplet.
\item
The gap exists, but the main part of the substance is accumulated by the
tail of the profile where it is possibly to use formally the AT due to
the previous note.
\end{itemize}

The first opportunity has conceptional significance and requires the special
consideration. It was completely done in \cite{PhysicaA}.  Here the
possibility
to grasp this situation is ensured by the universal character of the spectrum.

 The second opportunity is automatically accounted by the PT. The additive
character  here can be seen  if we take into account namely the integral
definition of $\beta_{eff}$. The integral character of the definition
of $\beta_{eff}$ instead of the level character of the definition of
$\beta_{st}$,
$\beta_{fin}$ allows to grasp the additive superposition of the long tails
of the profile.

The additive
character   of the superpositions of the profiles is the main feature
of the both limit situations described above.
As it is
stated in \cite{book} (page 315) the first kinetic model is the most natural
in the first situation. It is the corresponding model for the account
of the additive character. So, it have to be the corresponding model for the second
opportunity.

Despite the clear significance of the PT it is difficult for the
experimenters
to imagine the qualitative picture of the process and to compare the roles
of the previous AT and the new PT. So, here some simplifications will
be presented which will lead to more clear image of the process of
condensation.
One has to note that these simplifications are rather rough and ensure
only the write qualitative character of the process. Certainly one can
use directly the PT for the accurate results. The validity of  the PT
isn't violated below.


Note that the leading idea of consideration of the condensation process
is the avalanche character of kinetics. We shall use this property here
once more.

According to the avalanche character the essential error in the quantity
of the condensated substance leads to the rather small error in the number
of the droplets.  One can take here the total quantity of the substance
or the the substance condensated in the separate droplet.

Also due to the avalanche character of the process one can speak about
the characteristic size of the droplet, about the characteristic time
scale of the period of the droplets formation, about the back front of
the size spectrum.

We shall take for example the diffusion regime of the droplet growth. There
is no need to consider the transition to the free molecular regime because
it was completely discussed in \cite{PhysicaA}.

All definitions can be found in \cite{PhysicaA}.

\section{The length of relaxation}

>From the primitive formal point of
view the stationary profile leads to the divergence of the substance consumed
by the droplet (all the excess must be in the droplet) and for the
superposition
of the profiles from the other droplets. Really at the stationary profile
$n(\infty)  - n(r) \sim r^{-1}$ which can not be integrated. So, the substance
inside the droplet goes to $\infty$. On the other
hand the sum $\sum_i \frac{1}{
i}$ which  presents  the  superposition  of  the  profiles  of  the
droplets lying
on one line goes to $\infty$ and this can not be appropriated.

It seems that it closes the problem of the stationary profiles. But the
tails can be rather long and according to $\int_{1}^{\infty} \frac{1}{x}
dx \sim \ln(x) \rightarrow \infty$ can accumulate the main quantity of
the substance.

We shall start from the AT. The evolution of the elementary perturbation
can be described by the Green function of the free space
$$
G \sim exp(-\frac{r^2}{4Dt})
$$
In a very rough approximation (we are going to present the trivial
explanation)
one can see the plane region
$ r < \sqrt{4Dt}$ and the negligible tail $r> \sqrt{4Dt}$.

On the base of this consideration one can say that
$$
R_{rel} = \sqrt{4Dt}
$$
presents the characteristic length of relaxation.
One can take for $t$ the characteristic length of the period of the droplets
formation $t_1$. Then one can say that $r_{rel}$ can be treated as the
length of the stationary profile of the vapor density.

In the AT  for the finish of the droplets formation one can write
\begin{equation} \label{rrr}
N_{tot} \nu_1 = \Phi n_{\infty} / \Gamma ,
\end{equation}
where $\nu_1$ is the characteristic size of the droplets at the period
of the droplets formation.
Here $\Phi$ marks some characteristic value of the supersaturation during
the period of the droplets formation.

The mean distance between droplets can be estimated as
$$
R_{mean} = N_{tot}^{-1/3}
$$

For the rate of the droplet growth one can take the ordinary law
\cite{PhysicaA}
$$
\frac{d \nu}{dt} \sim
v_l^{1/2} (\Phi  n_{\infty} D )^{3/2} t^{1/2}
$$
and after the integration
$$
\nu
\sim
v_l^{1/2}       \Phi^{3/2} D^{3/2} n_{\infty}^{3/2} t^{3/2}
$$
Then
$$
\nu_1
\sim
v_l^{1/2}       \Phi^{3/2} D^{3/2} n_{\infty}^{3/2} t^{3/2}_1
$$

Then after the substitution one can come to
$$
Dt_1 =
\Gamma^{-2/3}
\Phi^{-1/3} (\frac{v_l}{v_v})^{-1/3} N_{tot}^{-2/3}
$$
where
$v_v = n_{\infty}^{-1}$
is the volume for the molecule in the saturated vapor.
This  leads to
$$
R_{rel} =
\Gamma^{-1/3} \Phi^{-1/6} (\frac{v_l}{v_v})^{-1/6} R_{mean}
$$
Introduce parameter
$$
\sigma = \Gamma^2 (\frac{v_l}{v_v}) \Phi
$$
One can see that parameter $\sigma$ can take the arbitrary values in
comparison
with $1$. The
inequality $\Gamma \gg 1$ can be compensated by the ordinary observed
inequality
$v_l \ll v_v$.

In the bubbles formation one has to substitute $v_v$ by $v_l$ and $v_l$
by $v_v$ ($\Phi$ here isn't important). Then one have to come to
$\sigma \gg 1$.

The final result will be
$$
R_{rel} = \sigma^{-1/6} R_{mean}
$$

 When $\sigma \ll 1$ one can use the stationary profile of density from
the point of view of AT which ignores the non-stationarity of the
tails of the profiles. The estimate
will be given in the next section.

It is rather natural to say that it is sufficient to have the stationary
profile
at the distances $R_{mean}$ (Certainly one  can  imagine  that  the
tails in
the superposition from the other droplets lying far (not neighbors)
are very important and they
are non stationary. The exclusion of this situation can be given analytically
and it will be directly seen later.

\section{The form of the stationary profile}

Now we see that it is
really interesting to investigate the stationary profiles
of the density.
The form of the profile is given by the very simple formula
$$
n_{st}(r) = n(\infty) - \frac{R_d}{r}[n(\infty) - n_{\infty} ]
$$
where
$R_d$ is the size of the droplet. It can be found by the
$$
R_d =
(\frac{3v_l \nu}{4 \pi})^{1/3}
$$

One can define now the boundary between the additive region and the
non-additive
region as the distance where
$$
\zeta(r) - \zeta(\infty) = \frac{\zeta(\infty)}{\Gamma}
$$

One can see that this boundary (let us denote it via $R_{add}$) is equal
to
$$
R_{add} = \Gamma R_d
$$

Define by $Q_{na}$ the quantity of the substance disappeared from the
non-additive region
\begin{equation} \label{qna}
Q_{na} =
\int_{R_d}^{R_{add}} [ n(\infty) - n(r) ] d^3 r
\end{equation}
One can easily calculate this integral in the approximation of the stationary
profile
$$
Q_{na} =
\int_{R_d}^{\Gamma R_d} \frac{R_d}{r} [n(\infty) - n_{\infty}] 4\pi r^2
dr =
\frac{3}{2} \sigma \nu
$$

>From the last expression it is seen that for $\sigma > 1$
$Q_{na} > \nu$ and the stationary approximation can not be valid for all
$r< R_{add}$.

Now it is clear that for $\sigma >1$ one can not use the
AT and has to use the PT. For $\sigma <1$ the question is open now.

Consider these questions from the point of view of the PT. Having taken
into account the mentioned representation for the Green function (certainly
it is worth correcting it by the normalizing factor and the intensity
of the vapor consumption) one can see by the analysis of the integrals
of the type  $\int r^2 \exp(-r^2/4Dt) dr$ that the essential part
of the substance is accumulated in the region with the stationary
profile.\footnote{One
has to calculate the Green function with the factor coming from the intensity
of the vapor consumption. The integral will be with another power of the
pre-exponential factor but the final conclusion about the essential part
will remain valid.}

When we speak about the essential part of the quantity
 we mean that this part is rather great and the ratio of this quantity
to the total quantity doesn't attain $0$ and $1$ or go to these values.
Due to the avalanche character of the process of  condensation  the
substitution
of the total quantity to the essential part doesn't lead to the great
variation of the kinetics of the process.

With the help of the last remark now one can give the qualitative description
of the profile. Until $r < R_{rel}$ the profile is close to the stationary
one. When $r > R_{rel}$ the profile isn't so essential.

Consider the situation $\sigma >1$.
It seems that for the validity of the stationary approach one needn't
to require that $R_{rel} \geq R_{mean}$ but only that $R_{rel} \geq \Gamma
R_d$. Really, all  tails  at  $r>\Gamma  R_d$  can  be  treated  as
corresponding
 to the additive
region.  Note, that there is absolutely no need to know the concrete form
of the profile in the additive region but only the  quantity condensated
in this region (which is the total quantity (the number of the molecules
in the droplet) except the quantity in the non-additive region (which
can be at $R_{rel} > \Gamma R_d$ easily calculated in the stationary
approximation)).

To show that there is no profit to consider $\Gamma R_d$ instead $R_{mean}$
compare these two values.

For the AT we can use the known equality
(\ref{rrr}) and come to
$$
\frac{R_{mean}}{\Gamma R_d} =
(\frac{3}{4\pi})^{1/3} \sigma^{-1/3}
$$
This shows that for $\sigma >1$ (when we have the problems with relaxation)
the new estimate can not also lead to the relaxation for $r<\Gamma R_d$
and there is no profit to use it.

As we have noted that the essential part of the condensated substance
is taken from the distances corresponding to the stationary profile of
the substance density we can introduce the distance $R_s$ by the condition
that
$$
\int_{R_d}^{R_s} n(\infty) - n_{st}(r) d^3 r = \nu
$$
Calculate this distance. Then
$$
R_s = (\frac{2}{3\sigma})^{1/2} \Gamma R_d
$$
or
$$
R_s = (2/3)^{1/2} (4 \pi / 3)^{1/3} \sigma^{-1/6} R_{mean}
$$
or
$$
R_s =
(2/3)^{1/2} (4 \pi / 3)^{1/3}  R_{rel}
$$
The absence of the parameters in the last equation is important and is
necessary for the validity of the statement that the essential part of
the  condensated substance comes from the distances with the stationary
profile.

\section{Situation $\sigma < 1$}

 In the opposite situation $\sigma > 1$ the AT has no perspectives because
even in
the non-additive region the stationary distribution is going to be
violated\footnote{
The stationary distribution will be violated at the end of the non-additive
region. This end will be attained due to the violation of the profile
which goes to be the small one but not when the excess of
stationary profile goes
to cross the level $n_{\infty} \zeta/ \Gamma$.}.

In the situation $\sigma <1$ the naive interpretation of the
boundary of the exhausted region  in the PT will lead to the big error.
Really at $\sigma <1$ the main quantity of the consumed substance is in the
additive region.
One can see this from expression (\ref{qna}) which can be used due to the
stationary character of the profile in the non-additive region in this
situation.

So, one has to use the integral definition of the boundary
of the exhausted region as it was done in \cite{PhysicaA} in the definition
of $\beta_{eff}$.
The reason of the integral style of the definition of $\beta_{eff}$ instead
of the level definition of $\beta{st}$ and $\beta_{fin}$ is to grasp the
situation with the essential long tails of the density profile.
But in the additive region there is no need to speak about the overlapping
of the exhausted regions. So, one has to use the first model as it was
done in \cite{book}, page 315.
The closeness of the results of all models guarantees the universal form
of the condensation kinetics.

Now we can show how to see the transition of the AT to the PT.
 One can explicitly calculate the effect from the gap in the stationary
approximation (which can be used when $\sigma < 1$) by the integration.
The remarkable fact is that
the quantity of the substance  $Q_{hna}$ accumulated from the region
$r<\Gamma R_d$ with the layer
$\Phi n_{\infty} / \Gamma$ (this comes from the formal
interpretation of the AT as the
theory with the exhauted regions) is comparable with $Q_{na}$
$$
Q_{hna} = \frac{2}{3} Q_{na}
$$
Then the asymptotical account of the gap can be attained by the
renormalization
of the quantity of the substance accumulated by one droplet
$$
Q \rightarrow Q(1-\frac{1}{2} (\frac{\Gamma R_d}{R_{mean}})^{1/3})
$$
This leads to the renormalization at $\sigma \ll1$
for the total number of the droplets
$$
N_{tot} \rightarrow  N_{tot} (1+\frac{2}{5} \frac{1}{2}
 (\frac{\Gamma R_d}{R_{mean}})^{1/3})
$$


\begin{thebibliography}{90}
\bibitem{PhysicaA}
V.Kurasov Physica A 226 (1996) 117-136

\bibitem{book}
V Kurasov,  Universality in kinetics of the first order phase transitions,
Chemistry Research Institute of St.Petersburg University, St.Petersburg,
1997, 400 p.

\end{thebibliography}
\end{document}